\documentstyle[12pt,aaspp4]{article}

\slugcomment{}

\lefthead{Wada and Norman}
\righthead{Multi-phase gas dynamics}

\newcommand{\beq}{\begin{equation} }
\newcommand{\eeq}{\end{equation}}
\newcommand{\bea}{\begin{eqnarray} }
\newcommand{\eea}{\end{eqnarray}}
\def\mbf#1{\mbox{\boldmath ${#1}$}}

\begin{document}
\title{The Global Structure and Evolution of a Self-Gravitating Multi-phase Interstellar Medium in a Galactic Disk}
\author{Keiichi Wada\altaffilmark{1} and Colin Norman\altaffilmark{2}}
\affil{Department of Physics and Astronomy, \\ Johns Hopkins University, Baltimore, MD 21218}


\altaffiltext{1}{National Astronomical Observatory, Mitaka, 181, Japan (email: wada@th.nao.ac.jp)}
\altaffiltext{2}{Space Telescope Science Institute, Baltimore, MD 21218 (email: norman@stsci.edu)}


\begin{abstract}
Using high resolution, two-dimensional hydrodynamical simulations, we
investigate the evolution of a self-gravitating multi-phase
interstellar medium in the central kiloparsec region of a galactic
disk.  We find that a gravitationally and thermally unstable disk
evolves, in a self-stabilizing manner, into a globally quasi-stable disk
that consists of cold ($T < 100$ K), dense clumps and filaments
surrounded by hot ($T > 10^4$ K), diffuse medium. The
quasi-stationary, filamentary structure of the cold gas is
remarkable. The hot gas, characterized by low-density holes and voids,
is produced by shock heating. The shocks derive
their energy from differential rotation and gravitational
perturbations due to the formation of cold dense clumps. In the
quasi-stable phase where cold and dense clouds are formed, the
effective stability parameter, $Q$, has a value in the range 2-5. The
dynamic range of our multi-phase calculations is $10^6 - 10^7 $ in
both density and temperature. Phase diagrams for this turbulent medium
are analyzed and discussed.
\end{abstract}




%

\section{INTRODUCTION}
\label{intro}
We model the multi-phase and inhomogeneous interstellar medium (ISM)
in the inner region of a galactic disk including fundamental physical
processes crucial for understanding star formation, global and local
dynamics of the ISM in galaxies, and aspects of galaxy formation such
as feedback. Most numerical simulations of the ISM and of star
formation in galaxies have assumed simpler ISM models, e.g. an
isothermal or nearly-isothermal equation of state, and either a smooth
medium or discrete clouds.  There are some notable
exceptions. Pioneering work by Bania \& Lyon (1980), in which the
effects of OB stars on the multi-phase ISM have been investigated
using two-dimensional hydrodynamics using 40$\times$40 grid cells for
a 180 pc$\times$180 pc region.  More recently, Rosen, Bregman \&
Norman (1993) and Rosen \& Bregman (1995) have studied dynamics of the
multi-phase ISM.  Their simulations are for two-fluids (gas and stars)
in two-dimensions, but they ignored the self-gravity of the gas and
the effect of galactic rotation.  Self-gravity of the gas is necessary
to produce very high density clouds which are the sites of star
formation. Rotation and the associated differential shear are also
important elements for the global gas dynamics and ISM structure.
V$\acute{\rm{a}}$zquez-Semadeni et al. (1995) and Passot et al. (1995)
have studied self-gravitating supersonic turbulence with 2-D
hydrodynamic and magnetohydrodynamic simulations. They also took into
account Coriolis force and large-scale shear. They assumed, however,
periodic-boundaries with a local-shearing coordinate.  They also
introduced an artificial mass diffusion term in the continuity
equation. As discussed in their paper, this term smoothes out the
density gradient, and prevents the generation of large density
contrasts and strong shocks. A full three-dimensional simulations of
gas and stars in a galaxy has been made by Gerritsen \& Icke (1997),
using the SPH technique.  They have a roughly two-phase structure of
the ISM.  Although the SPH is quite powerful for three-dimensional
problems, it is hard to achive a pc-scale resolution in 3-D for the
ISM in a whole galaxy.  One cannot avoid introducing {\it ad hoc}
assumptions for star formation and its energy feedback to the ISM
without a pc-scale resolution.

In this {\it Letter}, we report on the structure and global evolution
of multi-phase ISM models in two-dimensions, taking into account
self-gravity of the gas, galactic rotation, radiative cooling, and heating
due to UV background radiation. The effects of heating due to stellar
winds from massive stars and SNe are discussed in a subsequent paper.
We use an Eulerian hydro-code without periodic boundary
conditions.  Our code allow us to handle over seven orders of
magnitude for density and temperature in a $\sim$ kpc-scale region around the
galactic center with $\sim$ pc-scale resolution.

\section{NUMERICAL METHOD AND MODELS}
\setcounter{footnote}{0} We solve the following equations numerically
in two-dimensions to simulate the evolution of a rotating disk.

  \bea
\frac{\partial \rho}{\partial t} + \nabla \cdot (\rho \mbf{v}) &=& 0,
\label{eqn: rho} \\ \frac{\partial \mbf{v}}{\partial t} + (\mbf{v}
\cdot \nabla)\mbf{v} +\frac{\nabla p}{\rho} + \nabla \Phi_{\rm ext} +
\nabla \Phi_{\rm sg} &=& 0, \label{eqn: rhov}\\ \frac{\partial
E}{\partial t} + \frac{1}{\rho} \nabla \cdot ((\rho E+p)\mbf{v}) &=&
\Gamma_{\rm UV} - \rho \Lambda(T_g), \label{eqn: en}\\ \nabla^2
\Phi_{\rm sg} &=& 4 \pi G \rho, \label{eqn: poi} \eea

 where, $\rho,p,\mbf{v}$ are the density, pressure, and velocity of
the gas, and the specific total energy $E \equiv |\mbf{v}|^2/2+
p/(\gamma -1)\rho$, with $\gamma= 1.4$.  We assume a time-independent
external potential $\Phi_{\rm ext} \equiv = -(27/4)^{1/2}v_c^2/(R^2+
a^2)^{1/2}$, where $a$ is a core radius of the potential and $v_c$ is
the maximum rotational velocity.  We also assume a cooling function
$\Lambda(T_g) $ $(10 < T_g < 10^8 {\rm K})$ (\cite{SN}) and a heating
function $\Gamma_{\rm UV}$.\footnote{ In previous simulations on the
multi-phase ISM, the minimum temperature was assumed as 300 K
(\cite{RB93}; \cite{RB95}; \cite{VZ95a}; \cite{VZ95b}) or 100 K
(\cite{VZ96}).} We assume a uniform UV radiation field, which is
normalized to the local interstellar value, and photoelectric heating
of grains and PAHs.

The hydrodynamic part of the basic equations is solved by AUSM
 (Advection Upstream Splitting Method) (\cite{LS}) with a van
 Leer-type flux splitting process (Liou 1996).  After testing this
 code for various hydrodynamical 1-D and 2-D problems, we find that
 AUSM is as powerful a scheme for astrophysical problems as are the
 PPM (\cite{WC}) and Zeus (\cite{SM}) codes.  We achieve third-order
 spatial accuracy with MUSCL (\cite{VL}).  To satisfy the TVD
 condition using MUSCL, we use the minmod limiting function.  More
 details about our numerical code and test results are described in
 Wada \& Norman (1998, in preparation).

We use $1024^2$ Cartesian grid points covering a 2 kpc $\times$ 2 kpc
region. Therefore, the spatial resolution is 1.95 pc with $2048^2$
grid cells. A periodic Green function is used to calculate the
self-gravity for the $1024^2$ grid points (\cite{HE}).  The
second-order leap-frog method is used for the time integration.  We
adopt implicit time integration for the cooling term.

The initial condition are an axisymmetric and rotationally supported
disk with the Toomre stability parameter $Q = 1.2$ over the whole
disk.  Random density and temperature fluctuations are added to the
initial disk. These fluctuations are less than 1 \% of the unperturbed
values and have an approximately white noise distribution. The initial
temperature is set to $10^4$ K over the whole region. In ghost zones
at the boundaries, all physical quantities remain at their initial
values during the calculations. From test runs we found that these
boundary conditions are much better than `outflow' boundaries, because
the latter cause strong unphysical reflection of waves at the
boundaries.

%
\section{RESULTS}
%
\subsection{Morphology and Evolution}
Figure 1 shows the density and temperature distribution of the central
2 kpc $\times$ 2 kpc region at $t=166$ Myr.  The structure is
quasi-stable after $t\sim 50$ Myr.  The density range is from
$10^{-1}$ to $10^6 M_\odot$ pc$^{-2}$, which corresponds to about
$0.03$ to $3\times 10^5$ cm$^{-3}$.  The most prominent feature of the
resulting quasi-stable structure is its filamentary appearance. The
high density clumps are embedded in less dense filaments
($\Sigma_g \sim 10^{3-4} M_\odot$ pc$^{-2}$ and $T_g \lesssim 100$ K).
The characteristic size of the highest density clouds ($\Sigma_g \geq 10^5 M_\odot$
pc$^{-2}$) is about 5-50 pc.
It is notable that the density and temperature
structures are well correlated: high temperature gas corresponds to
low density gas, and low temperature gas corresponds to high density
gas.  This means that the thermal structure in this system is driven
by the density, because the radiative cooling is very effective.  The
temperature of the very low density gas in the voids sometimes reaches
$10^6$ K, due to shock heating. The energy source of this
heating is the turbulent, random motion of the gas, at velocities $\sim
100$ km s$^{-1}$.  The maximum vorticity is $\sim 40$ km s$^{-1}$
pc$^{-1}$.  There are strong local shear motions associated with the
filaments.

Figure 2 is a phase diagram ($p_{\rm th} \equiv \Sigma_g T_g$
vs. $\Sigma_g$) for the model shown in Fig.1.  The left and right
diagonal lines correspond to states in which the gas temperature
is $10^4$ and 10 K.  There are also two less prominent `ridges'
between the hot and cold phases, i.e. $(p_{\rm th}, \Sigma_g) =
(3,1)-(2.5,2)$ and $(2.5,2)-(4,4)$.  One ridge corresponds to
filaments with temperature, $T_g \sim 200$ K.  The second ridge
indicated by a dotted line 
has `negative $\gamma$' where $\gamma$ is the adiabatic index.  The
gas in this state is thermally unstable.  In spite of this instability
the gas component does not have temporal variations.  This is not a
paradox, if we consider the kinematic or turbulent pressure of the gas as
well as the thermal pressure.  As can be seen in this
phase diagram, the system cannot be described as a simple two- or
three-phase media with pressure equilibrium between the phases.

Although the initial $Q$ value is 1.2 (i.e. the disk is stable for an
axisymmetric modes), the effective $Q$ value becomes much less than
unity after a few Myr (see \S 3.2 and Fig.3), because the cooling time
is very short.  The gas temperature decreases from the initial
value ($\sim 10^4$ K) to an equilibrium temperature ($\sim 10^2$ K)
within $10^5$ yr.  Since the initial density is higher in the central
region of the disk than in the outer region, gravitational
instabilities begin in the inner few 100 pc and develop outward.  At
$t \sim 10$ Myr ($\sim$ one rotational period at $R\sim 200$ pc), the
gravitational instabilities in the central 500 pc are already in a
non-linear phase, where the maximum density contrast is about $10^6$.
At $t\sim 25$ Myr, the instability grows non-linearly over the whole
disk, and many clumps, filaments and low density voids are formed.
The outer region of this multi-phase disk expands until, at about
$t=50$ Myr, it reaches a quasi-stable state.  The total
entropy of the system increases very rapidly in the first 20 Myr, and
it remains roughly constant after 30 Myr. The entropy generation by
shocks is balanced by radiative cooling to maintain this stationary
state.

About 80 \% of the mass is in the cold ($T_g<100$ K) phase.  In other
words, the cold, dense clumps dominate the total gas mass.  The second
most dominant phase is the warm ($100 <T_g <9000$ K) gas.  The mass of
this warm gas increases from $10^7 M_\odot$ to $10^8 M_\odot$ in the
first 10 Myr.  It is notable that the gas mass in the hot phase ($T_g
>10^4$ K) increases by a factor of order $\sim 10^2$ for the first 30
Myr. In the quasi-stable phase, the hot gas occupies about 0.3\% of
the total mass.  As expected, on the other hand, the volume filling
factors of the hot, warm and cold gas indicate that the hot gas ($T_g
\geq 9000$ K) occupies a volume $10^{2.5}$ times larger than the cold
gas. The volume filling factor for the gas in the temperature range
$11000 \leq T_g < 10^5$ K increases from 0 to 30\% until $t=30$
Myr. In contrast, the cold gas fraction decreases from 48\% to 10\% in
the same period.  After 30 Myr, the volume filling factor of each
component does not change appreciably, and this also shows that the
system reaches a quasi- equilibrium state.

The radial mass distribution has roughly exponential shape after 50
Myr, and the profile at $165$ Myr becomes a steeper exponential.  The
density inside $R=100$ pc becomes about 4 times larger at $t=$ 165
Myr.  The total gas mass inside this region is about $3\times 10^7
M_\odot$, which is comparable to the dynamical mass.  The evolution of
the radial profile implies outward transport of angular momentum. In
fact, 5 \% of total specific angular momentum is transfered through
the outer boundary. This is due to random motion of the flow caused by
the non-linear evolution of the gravitational instability.  Although
the global density profile is exponential, it fluctuates by a factor
of $\sim$ 2-3 in over scales of order $\sim$100 pc.  These
fluctuations correspond to complexes of clouds and filaments that form
weak spiral-like patterns as seen in Fig.1.

\subsection{Self-stabilization due to Dynamical Heating}

The evolution of Toomre's Q parameter in this system is very
interesting, because this parameter has been often supposed to be a
good criterion for star formation in disk galaxies (e.g. \cite{KN89}).
Figure 3 shows a radial distribution of an effective $Q$ value
($Q_{\rm eff}$) at $t=$ 10, 25, 50, 166 Myr of a model without star
formation.  $Q_{\rm eff}$ is defined as $ Q_{\rm eff}(R) \equiv
{\kappa \sigma_v}/{\pi G \Sigma_g}, $ where $\kappa$ is the epicyclic
frequency.  The azimuthally averaged ($\Delta R =20$ pc) velocity
dispersion $\sigma_v$ is defined by $\sigma_v \equiv \sqrt{\langle
v^2\rangle_{\Delta R} - \langle v \rangle^2_{\Delta R} }$ with $v
\equiv |\mbf{v}| +c_s$, where $\mbf{v}$ the flow velocity and $c_s$ is
the sound velocity of the gas.  As mentioned above, the initial
$Q_{\rm eff}$ becomes far less than unity almost everywhere in the
disk after 1 Myr.  However, after 50 Myrs, $Q_{\rm eff}$ increases
above the marginal stability value for the whole disk.  This means
that the system is globally self-stabilized.  We found that the
stabilization is mostly due to the increase of the velocity dispersion
of the gas, rather than large sound velocity in the high temperature
gas arising from the shock heating.  The averaged velocity dispersion
increases to $\sim$ 15-20 km s$^{-1}$ in 50 Myr, whereas the averaged
sound speed increases from about 3 km s$^{-1}$ to 8 km s$^{-1}$.
\footnote{It is of interest to compare this velocity dispersion with
the internal velocity dispersion of GMCs to understand the mechanism
that maintains the interal motion of GMCs. Although the dispersion,
$\sim$ 15-20 km s$^{-1}$, is in the same order of the observed line
width for GMCs in the central region of our Galaxy, our spatial
resolution ($\sim 2$ pc) is still insufficient for a detailed
comparison with the internal motion of GMCs. }

The multi-phase disk is {\it locally} unstable, however it is {\it
globally} stable, as shown by $Q_{\rm eff}$. This suggests that the
Toomre's $Q$ might not be useful directly here as a star formation
criterion, because there are many cold and dense clumps formed even at
effective values of Q significantly greater than unity. In such cold
and dense clumps, the Jeans length is much smaller than the
characteristic size of the clumps. Therefore, we expect star formation
in the clumps. Energy feedback from such stars to the ISM, the
ultimate fate of the dense clouds and the global structure of the
multi-phase ISM are interesting problems and are discussed in
subsequent papers.

It is notable that the multi-phase, quasi-stable ISM is formed from a
highly gravitationally and thermally unstable state without heating
due to SNe. The diffuse UV heating does not contribute to the global
stabilization. Shock heating is a dominant heating mechanism for
$T_g> 10^4$ K in this system. Shearing of the disk induced by
differential rotation and gravitational perturbation from clumps are
the main energy source for the shock heating.

%
\section{SUMMARY AND DISCUSSIONS}
%
We have shown that a globally stable, multi-phase ISM is formed as a
natural consequence of the non-linear evolution of a self-gravitating
gas disk. The gas is in so many and such various phases represented by
a wide range of density and temperature that multi-phase is probably
an inadequate description.  The gas has properties more like a phase
continuum.  The density ranges over seven orders of magnitude from
$10^{-1}-10^{6} M_{\odot}$ pc$^{-2}$, and the temperature extends from
$10- 10^8$ K.  We found that the ISM cannot be expressed simply by the
two-phase or three-phase model (e.g. \cite{MO}; \cite{IH};
\cite{NI89}; \cite{NF96}), because the gas dynamic processes, such as
turbulence, global rotation, shear motion and shocks, are at least as
important as thermal processes.  Our high resolution Eulerian
hydrodynamic code allows us to handle gas dynamics on a kiloparsec
scale with a few parsec resolution regardless of the gas
density. Possible star forming sites where the gas density is very
high ($n \gtrsim 10^4$ cm$^{-3}$) and the temperature is less than 100
K can be identified clearly. The evolution of supernova remnants in an
inhomogeneous and rotating media is fully followed as discussed in a
subsequent paper.  The multi-phase structure of the gas is very
complicated, and is characterized by filamentary and clumpy
substructure as well as low density holes and voids surrounded by the
denser media.

Although our numerical simulations show the potential of more
realistic models of the ISM and of star formation than in previous
simulations with lower resolution, there are a number of physical
aspects of the real ISM that should be considered in future models:
the ionization state of the gas; the non-uniform UV radiation field
due to massive stars; magnetic fields; the chemical evolution and
dynamics of the stellar component and its interaction with the gaseous
component.  Besides these physical processes, three-dimensional
simulations need to be considered. Our numerical scheme can be
extended to three-dimensions, and in fact such simulations are
currently possible but at the expense of either resolution or dynamic
range in the disk.

Our results show that the description of the ISM can be moved to a
more realistic level using techniques such as we have discussed here.
This has interesting implications for many aspect of galaxies.  In
particular, we should now consider the complex nature and dynamics of
the multi-phase and inhomogeneous ISM in theories of star formation,
galaxy formation and galaxy evolution.

\acknowledgments We would like to thank the referee, V. Icke, and
T. Hasegawa for their helpful suggestions.  We are grateful to
M. Spaans for providing us the cooling functions and stimulating and
useful discussions. We thank D. Strickland for his fruitful comments
on the draft. We also acknowledge stimulating discussions with the
members of the {\it HOT TOPICS} group at JHU.  KW thanks Yamada
Science Foundation for their support at STScI.  Numerical computations
were carried out on VPP300/16R at the Astronomical Data Analysis
Center of the National Astronomical Observatory, Japan.  This work has
been supported in part by NASA Grant NAG8-1133.

\clearpage

\newpage
\figcaption{Density (left) and temperature (right) map at $t=166$ Myr.
 The color bar is log-scale. The density range is $10^{-1}$ to $10^5 M_\odot$ pc$^{-2}$,
and temperature range is $10-10^8$ K.}
\figcaption{Pressure ($\Sigma_g T_g$) vs. density ($\Sigma_g$) phase
diagram for the model shown in Fig. 1. Three diagonal lines
show the gaseous temperature $10^4$ K, $190$ K, and $10$ K.
The dotted line means thermally unstable gas.}
\figcaption{Evolution of the effective Toomre Q as a function of radius.
Thin line with $+$, dashed line, dotted line, and solid line are 
for $t=$10, 25, 50, 166 Myr respectively. $Q_{\rm eff} <1 $ means the system isunstable.}

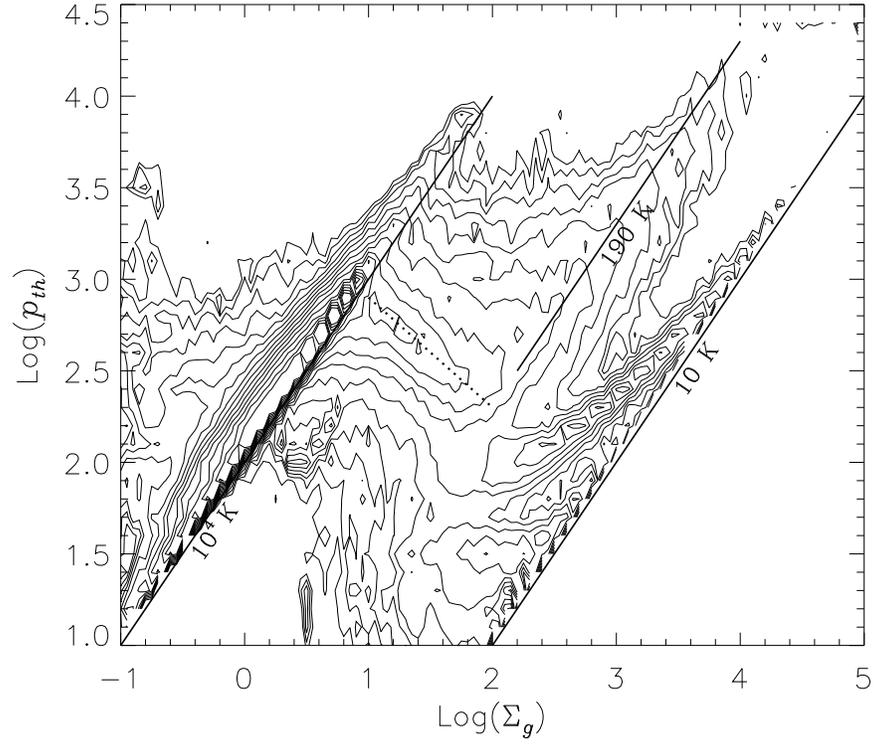
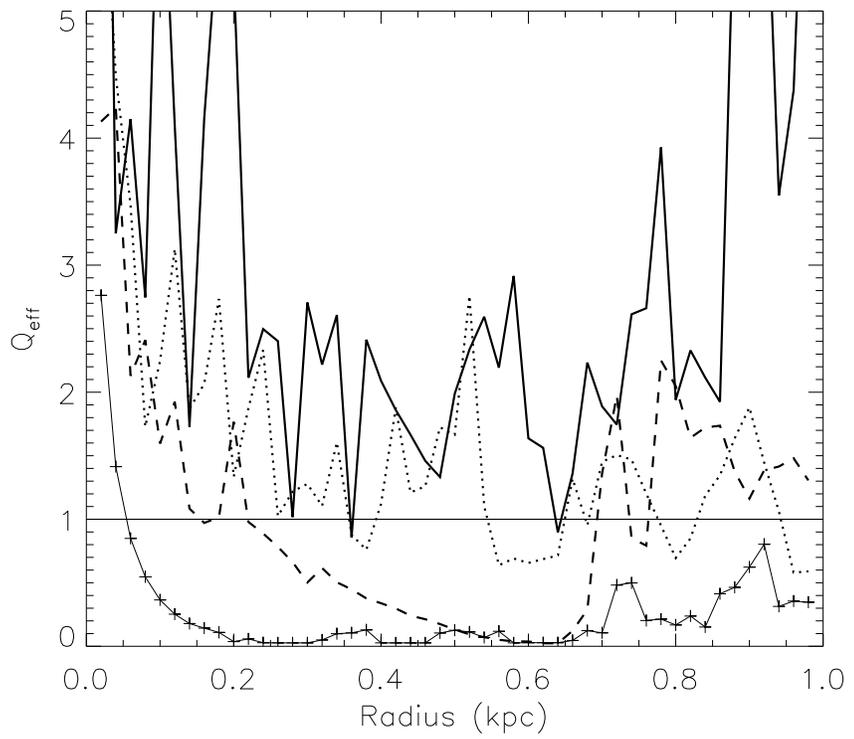
\end{document}